\documentclass[a4paper,12pt]{article}
\pdfoutput=1
\usepackage{amsmath,amsfonts,amssymb,mathrsfs,graphicx}
\usepackage[squaren]{SIunits}
\usepackage{natbib}

\allowdisplaybreaks[4] % for page break

\begin{document}

\newcommand{\ie}{{\it i.e.}}
\newcommand{\Ie}{{\it I.e.}}
\newcommand{\eg}{{\it e.g.}}
\newcommand{\Eg}{{\it E.g.}}
\newcommand{\cf}{{\it cf.}}
\newcommand{\etc}{{\it etc.}}
\newcommand{\eq}{Eq.}
\newcommand{\eqs}{Eqs.}
\newcommand{\Eq}{Eq.}
\newcommand{\Eqs}{Eqs.}
\newcommand{\Def}{Definition}
\newcommand{\fig}{Fig.}
\newcommand{\Fig}{Fig.}
\newcommand{\figs}{Figures}
\newcommand{\Figs}{Figures}
\newcommand{\Ref}{Ref.}
\newcommand{\Refs}{Refs.}
\newcommand{\Sec}{Section}
\newcommand{\Secs}{Sections}
\newcommand{\App}{Appendix}
\newcommand{\Apps}{Appendices}
\newcommand{\Tab}{Tab.}
\newcommand{\Tabs}{Tabs.}

% WW editing:
\newcommand{\stheta}{\sin^22\theta_{13}}
\newcommand{\deltacp}{\delta_\mathrm{CP}}
\newcommand{\ldm}{\Delta m_{31}^2}
\newcommand{\sdm}{\Delta m_{21}^2}
\newcommand{\equ}[1]{\eq~(\ref{equ:#1})}
\newcommand{\figu}[1]{\fig~\ref{fig:#1}}
\newcommand{\tabl}[1]{\Tab~\ref{tab:#1}}
\newcommand{\bi}{\begin{itemize}}
\newcommand{\ei}{\end{itemize}}
\newcommand{\ra}{\rightarrow}

%%%%%%%%%%%%%%%%%%%%%%%%%%%%%%%%%%%%%%%%%%%%%%%%%%%%%%%%%%%%%%%%%%%%%
%%%%                     Title-page                              %%%%
%%%%%%%%%%%%%%%%%%%%%%%%%%%%%%%%%%%%%%%%%%%%%%%%%%%%%%%%%%%%%%%%%%%%%
\begin{titlepage}

\renewcommand{\thefootnote}{\alph{footnote}}

\vspace*{-3.cm}
\begin{flushright}
% Report numbers

%IFIC/12-50
\end{flushright}

%\vspace*{0.2cm}

\renewcommand{\thefootnote}{\fnsymbol{footnote}}
\setcounter{footnote}{-1}

{\begin{center}
{\large\bf
Estimation of the neutrino flux and resulting constraints on hadronic
emission models for Cyg~X-3 using \textit{AGILE} data
} \end{center}}
\renewcommand{\thefootnote}{\alph{footnote}}

\vspace*{.3cm}
% \vspace*{.3cm}
{\begin{center}
{\large \today}
\end{center}
}

{\begin{center} {
		 \large{\sc
                 P.~Baerwald\footnote[1]{\makebox[1.cm]{Email:}
                 philipp.baerwalds@physik.uni-wuerzburg.de}} and
                 \large{\sc
		 D.~Guetta\footnote[2]{\makebox[1.cm]{Email:}
                 dafne.guetta@oa-roma.inaf.it}${}^,$\footnotemark[3]}
		 }
\end{center}
}
\vspace*{0cm}
{\it
\begin{center}

\footnotemark[1]
       Institut f{\"u}r Theoretische Physik und Astrophysik, 
       Universit{\"a}t W{\"u}rzburg, \\
       D-97074 W{\"u}rzburg, Germany \\
\footnotemark[2]
       Osservatorio astronomico di Roma,
       v. Frascati 33, \\
       I-00040 Monte Porzio Catone, Italy \\
\footnotemark[3]
       Department of Physics and Optical Engineering,
       ORT Braude, P.O. Box 78, \\
       Carmiel, Israel

\end{center}}

\vspace*{0.5cm}

{\Large \bf
\begin{center} Abstract \end{center}  }

In this work we give an estimate of the neutrino flux that can be expected from the microquasar Cyg~X-3. We calculate the muon neutrino flux expected here on Earth as well as the corresponding number of neutrino events in the IceCube telescope based on the so-called hypersoft X-ray state of Cyg~X-3. If the average emission from Cyg~X-3 over a period of 5 yr were as high as during the used X-ray state, a total of 0.8 events should be observed by the full IceCube telescope. We also show that this conclusion holds by a factor of a few when we consider the other measured X-ray states. Using the correlation of \textit{AGILE} data on the flaring episodes in 2009 June and July to the hypersoft X-ray state we calculate that the upper limits on the neutrino flux given by IceCube are starting to constrain the hadronic models, which have been introduced to interpret the high-energy emission detected by \textit{AGILE}.

\vspace*{.5cm}

\end{titlepage}

%\maketitle

\section{Introduction} 

\label{sec:intro}

One of the fundamental questions in astrophysics is if hadrons are 
present in the jets of astrophysical sources like microquasars (MQs),
quasars, and gamma-ray bursts (GRBs).
This is still an open issue not solved by the detection of photons or 
high-energy protons (who lose information on the originating source 
on their way from the source to us). 
Unlike high-energy photons and protons, neutrinos can travel cosmological distances
 without being absorbed or deflected. Therefore, neutrinos can provide information 
on astrophysical sources that cannot be obtained with high-energy photons and charged particles.
 However the weak interaction
 of neutrinos with matter also implies that they are very difficult to detect,
 requiring detectors with an instrumented volume of $\mathcal{O}(\kilo\meter\cubed)$. 
IceCube, completed in 2010 December, is the first kilometer-scale neutrino detector.
IceCube analyses include model-independent searches for a diffuse flux and searches for point sources.
 Recent efforts to detect higher energy neutrinos from sources outside our solar system 
yield important constraints on point sources and diffuse fluxes of possible sources in
the Galaxy (\cite{IceCube:2012sj}).

In this paper, we concentrate on one kind of astrophysical sources with relativistic jet,  MQs. 
These are Galactic X-ray binary systems, which
exhibit relativistic radio jets (\cite{Mirabel:1999fy,2001AIPC..558..221F}). These
systems are believed to consist of a compact object, a neutron
star or a black hole, and a giant star companion. Mass transfer
from the giant star to the compact object through the formation of
an accretion disk and the presence of the jets make them similar
to small quasars, hence their name ``microquasars.''
The observed radiation from MQ jets, typically is
in the radio and in some cases also in the IR band and it is consistent
with non-thermal synchrotron radiation emitted by a population of
relativistic, shock-accelerated electrons.
Recently high-energy emission, $\sim \giga\electronvolt$, has been detected from MQ
by \textit{AGILE} and \textit{Fermi}.

MQs can also be sources of $1$--$100 \, \tera\electronvolt$ neutrinos (\cite{Levinson:2001as}) that come
from the interaction of $\sim 10^{16} \, \electronvolt$ protons with
synchrotron photons emitted by the shock-accelerated electrons. 
The predicted fluxes should be detectable by large, km$^2$-scale effective area,
high-energy neutrino telescopes, such as the operating south pole
detector IceCube, see \cite{Ahrens:2002dv}. \cite{Distefano:2002qw} have extended
the work of \cite{Levinson:2001as} and predicted the neutrino flux for several MQs
whose radio data were available in the literature. They found that 
the largest number of neutrinos from bursting MQs is expected to come from Cyg X-3 and XTE J1118 +480.
 
In this paper, we want to study the role of the hadronic
component in the MQ jet on the high gamma-ray emission mechanism.
In order to perform this analysis, we use the IceCube and \textit{AGILE} data.

The \textit{AGILE} team has made a detailed study of
$\gamma$-ray emission from Cygnus~X-3 with the \textit{AGILE} satellite. 
The \textit{AGILE} discovery of transient
$\gamma$-ray emission from Cygnus~X-3 in 2008 April associated with a specific spectral state preceding a major radio jet ejection opened a new window of investigation of MQs.
Several other major $\gamma$-ray emission episodes from Cygnus X-3
have been detected by \textit{AGILE} and \textit{Fermi} since 2008 (\cite{Tavani:2009pm,2009Sci...326.1512F}).
The \textit{AGILE} discovered several transient
$\gamma$-ray emission episodes from Cygnus X-3 in the energy range $100 \, \mega\electronvolt$ -- $50 \, \giga\electronvolt$ 
during the periods 2009 June--July and 2009 December--2010 mid-June.
They found that the $\gamma$-ray emission from Cygnus X-3 is detectable by \textit{AGILE} not only during
relatively short (1--2 days) flares as in \cite{Tavani:2009pm}, but also during extended
periods lasting several days or weeks (as during 2009 June--July; \cite{Bulgarelli:2011qk}). Detecting continuous $\gamma$-ray emission during ``active'' phases is of great theoretical relevance for the modeling of Cygnus X-3. 
 
The main mechanism at the origin of the high-energy emission detected by \textit{AGILE} can be the inverse Compton (IC) scattering of high-energy electrons accelerated in the source with low-energy photons emitted by synchrotron process (\cite{2010MNRAS.404L..55D}).
However, another possibility, if hadrons are present in the jet, is that the $100 \, \mega\electronvolt$--$50 \, \giga\electronvolt$ emission is due to $\pi^0$ decay into $\tera\electronvolt$ $\gamma$-rays that trigger a cascade to GeV gamma rays. These $\pi^0$ may be produced in the jet by photohadronic interactions or by proton--proton collisions between protons in the jet and protons in the gaseous surroundings provided by the WR companion mass outflow (\cite{Piano:2012my,Romero:2003td}).\footnote{Note that it has been pointed out in \cite{2011A&A...528L...2T} that the approach from \cite{Romero:2003td} has a mistake in the boost factors, which needs to be corrected for accurate calculations.}

In this paper, we use a simplified jet model to estimate the number of neutrinos expected from Cyg~X-3 based on the X-ray data taken in 2009 during an episode of flares. 
In Section~\ref{sec:model}, we describe the simplified model we use for the jet of the
MQ and the photohadronic interactions happening inside of it. In
Section~\ref{sec:obsdata}, we derive the expected photon and proton densities inside the jet
of Cyg~X-3, which are needed for the simulation of the photohadronic
interactions, from the observed X-ray data. In
Section~\ref{sec:prediction}, we then use the derived densities to estimate the expected neutrino
number both in the case that the emission is due to IC and in the case that is
due to $\pi^0$ decay into $\gamma$-ray photons. In Section~\ref{sec:discussion}, we discuss our
results.

\section{A simplified jet model} \label{sec:model}

In this section, we want to describe our simplified jet model. The model we want
to apply to the MQ Cyg~X-3 was originally used for GRB which are also
assumed to originate from emitting jets. The main simplification is that we
totally neglect the dynamic of the propagating jet and assume that the outflow
is static. The high-energy protons as well as the X-ray and $\gamma$-ray photons are assumed to
be present in one interaction zone and emission will only originate from this
one emission region. Both the proton as well as the photon density are assumed
to be isotropic in the plasma rest frame of the jet. These assumptions are
essential for us to use the photohadronic interaction code based on
\cite{Hummer:2010vx} (model Sim-B), which is an analytical
parameterization based on SOPHIA, see \cite{Mucke:1999yb}. Starting from the
particle densities, the code calculates the result of the photohadronic
interactions leading to secondary particles which subsequently decay into
neutrinos. It is based on an analytical parameterization of the full
photohadronic interaction cross section and includes features, such as
individual treatment of secondary particles, helicity-dependent muon
decay, as well as neutrino flavor mixing, for details see \cite{Hummer:2010ai}. 
With this treatment we also incorporate the magnetic field effects discussed by \cite{Reynoso:2008gs}, which do play a role due to the assumption of neutrino production in the internal shocks.

We assume in concordance with \cite{Levinson:2001as} that the photon energy density inside the jet $U'_{\gamma}$ can be calculated from the emitted luminosity $L$ of the source by considering the emission passing the surface of the jet
\begin{equation}
	U'_{\gamma} = \frac{L}{4\pi \, \left( l^2 \cdot \sin^2 (\theta_j) \right) \, c \, \mathcal{D}^2}
	\label{equ:luminosity}
\end{equation}
with $l$ being the distance of the emission region from the central object, $\mathcal{D}$ being the Doppler factor of the jet, and $\theta_j$ being the opening angle of the jet. Note that the boost with the Doppler factor is needed, since $U'_{\gamma}$ is the energy density inside the shock frame. Additionally, we know that the photon energy density can be defined as
\begin{equation}
 	U'_{\gamma} = \int E'_{\gamma} \, n'_{\gamma} \, \mathrm{d}E'_{\gamma} \quad ,	
 	\label{equ:energydensity}
\end{equation}
with $n'_{\gamma}$ being the photon spectrum in particles per volume and per energy. For our photohadronic interaction calculation we will need $n'_{\gamma}$. The shape of this spectrum can be obtained from the photon data, while the normalization can be obtained using \equ{energydensity}.

Hence, by knowing the observed (energy integrated) photon flux $F_\gamma$ we can calculate the (isotropic equivalent) luminosity $L = 4\pi \, D_L^2 \, F_\gamma$, and subsequently the photon particle density inside the jet. Moreover, it is possible to calculate the energy carried by the magnetic field $B$ as well as the energy in protons from energy (equi-)partition arguments. These theoretical considerations need to be applied since we currently do not have the possibility to estimate the magnetic field strength by other means. Following the considerations of \cite{Levinson:2001as}, it is possible to estimate the magnetic field $B'$ (in G) by
\begin{equation}
	B' = \sqrt{8\pi \, \frac{\varepsilon_B}{\varepsilon_e} \, U'_\gamma} \label{equ:magneticfield}
\end{equation}
with $\varepsilon_B$ being the fraction of the jets total energy carried by magnetic field, $\varepsilon_e$ being the fraction carried by electrons/photons, and $U'_{\gamma}$ from \equ{luminosity} in $\text{erg} \, \reciprocal\second$. Accordingly, we can estimate that the energy density carried by protons is
\begin{equation}
 	U'_p = \frac{\varepsilon_p}{\varepsilon_e} \cdot U'_{\gamma}
	\label{equ:protonenergydensity}
\end{equation}
with $\varepsilon_p$ being the fraction of the jets total energy carried by
protons. The fractions, $\varepsilon_B$, $\varepsilon_e$, and $\varepsilon_p$, are estimated to be of the order of $0.1$,
see \cite{Distefano:2002qw} and \cite{Levinson:2001as}. Moreover, we can connect the proton spectrum
$n'_p$ to the energy density in protons $U'_p$ by adapting \equ{energydensity} for
protons. However, opposite to the photon spectrum, we do not have any direct
observational information on the shape of the proton spectrum. From
considerations on the Fermi acceleration we know that the spectrum should have a
form  $\propto E^{-\alpha_p}$ with $\alpha_p \simeq 2$. Furthermore, we know
that particles can only be accelerated to finite energies and a cut-off at a
critical energy $E_c$ should be expected.\footnote{The cut off we assume for our
calculations is of the form $\exp\left(-(E/E_c)^2\right)$; however, the actual
shape of the cut off is not relevant, only the position given by critical
energy, $E_c$, is.} We can estimate the critical proton energy by basic
considerations on the jet properties. 
In general, the critical energy of the protons is obtained by comparing the acceleration rate to the (total) loss rate. This total loss rate can have several contributions, such as synchrotron losses, adiabatic losses, losses from $p\gamma$- or $pp$-interactions, which have different energy dependencies. It is therefore needed to evaluate the maximal loss rate on a energy-dependent basis, and with respect to used parameters, e.g., a higher proton density would lead to higher $pp$-losses, while a higher photon density would to the same for $p\gamma$-losses. For the parameters we consider in this paper we have however found that there should be the following two scenarios:
a synchrotron limited case and a escape limited case. In the
first case, the size of the acceleration region is sufficiently large to not
affect the particle acceleration and synchrotron losses are considered to be the dominant contribution to the (total) energy loss rate. By comparing the acceleration time to the synchrotron loss time
it is possible to obtain the critical energy (in the plasma rest frame)
\begin{equation}
 	E'_{c1} = 2.01 \cdot 10^{11} \, \sqrt{\eta} \, \left(
\frac{B'}{1 \, \text{G}} \right)^{-\frac{1}{2}} \, \giga\electronvolt
	\label{equ:critenergy1}
\end{equation}
with $\eta$ being the acceleration efficiency and $B'$ being the magnetic field in the
plasma rest frame. For all the calculations in this paper we have assumed
$\eta = 0.1$. In the second case, the energy of the protons is limited by the
maximal energy the particles can reach before they escape the source.
Effectively, this is equivalent to comparing the size of the region to the Larmor
radius of the particles. Hence, this critical energy can be estimated by
comparing the acceleration time to the escape time from the jet which leads to
\begin{equation}
 	E'_{c2} = 3 \cdot 10^9 \, \eta \, \left( \frac{l}{10^{14} \,
\centi\meter} \right) \cdot \sin(\theta_j) \, \left( \frac{B'}{1 \, \text{G}}
\right) \, \giga\electronvolt \quad .
	\label{equ:critenergy2}
\end{equation}
The smaller of the two critical energies is the relevant critical energy for a given set of parameters.

\section{Observational data on Cyg~X-3} \label{sec:obsdata}

The information on Cyg~X-3 is comparably well documented and there already have been earlier studies on the neutrino emission of this source, such as \cite{Distefano:2002qw}. 
For our study we adopt the following parameters for the source:
\begin{eqnarray*}
	D_L &=& 7.2 \, \text{kpc} \quad ,  \\ %\simeq 2.222 \cdot 10^{22} \, \centi\meter 
	\mathcal{D} &=& 2.74 \quad ,   \\
	\theta_j &=&12\degree \quad ,  \\
	l &=& 10^{8} \, \centi\meter \quad .  
\end{eqnarray*}
$D_L$ is the distance between Earth and the source, with the result taken from \cite{Ling:2009kc}. The value of the Doppler factor $\mathcal{D}$ is the result of an assumed Lorentz factor $\Gamma = 1.70$ and a viewing angle $\theta = 14\degree$, taken from \cite{Distefano:2002qw} and \cite{Mioduszewski:2001ev}. The jet opening angle $\theta_j$ is also taken from \cite{Mioduszewski:2001ev}. The value for the radius of the emission region $l$ is set to the assumed collision radius of internal shocks in an MQ, as in \cite{Levinson:2001as}.
Note, however, that this assumed collision radius may actually be smaller or larger as there are currently no spectroscopic measurements at these wavelengths, as opposed to radio measurements. For this reason, the actual emission radius may be different and will introduce an uncertainty on the neutrino flux.

In a second step we need to take into account the kinematics of the photohadronic interactions to estimate which energy range of photons can act as the target photons for our highest energy protons. Assuming a maximal proton energy of $10^8 \, \giga\electronvolt$, we can estimate that X-ray photons with several $\kilo\electronvolt$ energy are needed for producing the $\Delta$-resonance. For this reason, we will use X-ray data from \cite{Koljonen:2010wa} as our target photons. Since we are interested in the phase directly before a flare we will use data for the hypersoft state from said paper. The used data ranges from $3.5$ up to $102 \, \kilo\electronvolt$, with the integrated luminosity we obtain being $L_{\gamma} = 3.77 \cdot 10^{37} \, \text{erg} \, \reciprocal\second$. With the help of \Eqs~\eqref{equ:luminosity} and \eqref{equ:energydensity} we can normalize our target photon density. Furthermore, we can estimate that for the given $L_\gamma$ the resulting magnetic field from \equ{magneticfield} is $B' = 8.8 \cdot 10^5 \, \text{G}$, while the resulting maximal proton energy is $E'_{p,c} = 5.5 \cdot 10^7 \, \giga\electronvolt$, given by \equ{critenergy2}. Hence, we can normalize the proton density as well through \equ{protonenergydensity}. The density, which we obtain using this method, corresponds to an injection rate of $2.5 \cdot 10^{23} \, \text{protons}\, \centi\meter\rpsquared \, \reciprocal\second$. Additionally, we use the same approach on the five other identified flux states from \cite{Koljonen:2010wa} to get an idea how the dynamic of the jet affects the neutrino flux prediction.

\section{Expected neutrino flux} \label{sec:prediction}

\begin{figure}
	\centering
	\includegraphics[width=0.5\textwidth]{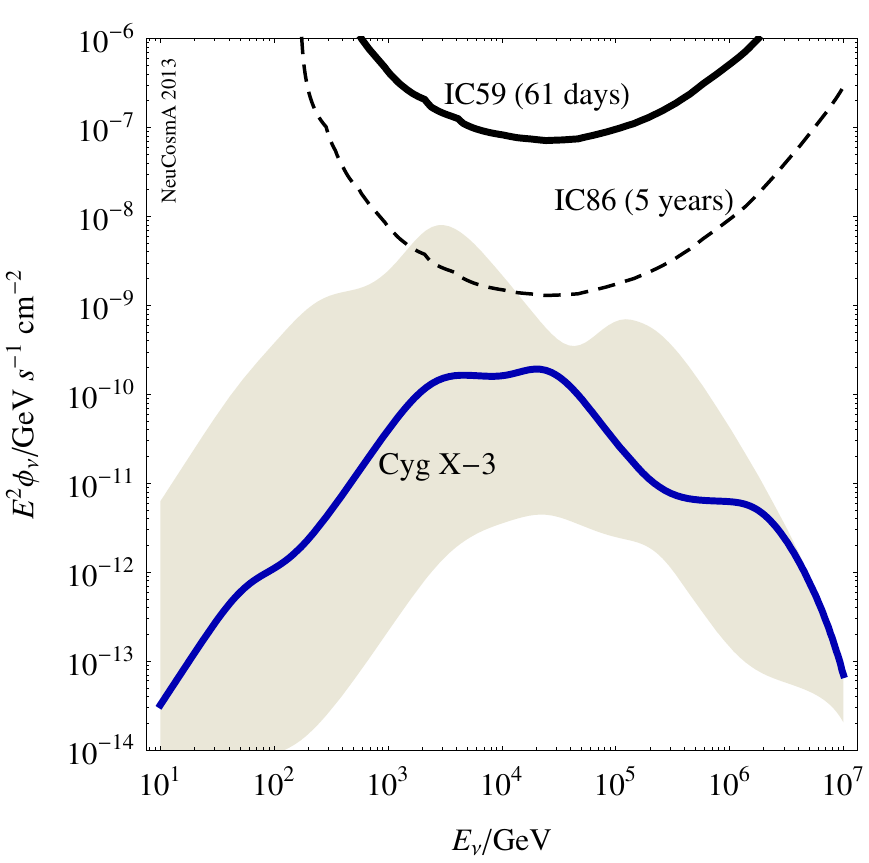} 
	\caption{In this plot, we show the expected neutrino flux from Cyg~X-3 derived for a simplified model of the microquasar jet, based on hypersoft X-ray data from \cite{Koljonen:2010wa} (gray solid curve). The gray shaded region represents the uncertainty due to the not directly measured emission radius by varying the collision radius $l$ by one order of magnitude around $10^8 \, \centi\meter$. For comparison, we give the (extrapolated) differential flux limits for IC59 during the 61 days of the flaring episodes (solid black curve) as well as an extrapolation of the expected one for IC86 (full detector) after 5 yr of data taking (thin dashed curve). As one can see both do not reach the flux of this single microquasar, even though the full detector could with a longer exposure. The differential limits are based on the solid-angle-averaged effective areas at final cut level of the time-integrated point-like source search from \cite{Abbasi:2010rd}.}
	\label{fig:CygX-3neutrinos}
\end{figure}

With the normalized spectra we obtained in the previous section we can
calculate the shape of the neutrino flux at Earth after flavor mixing.
Especially, the shape of the muon neutrino flux is relevant for the current
neutrino experiments such as IceCube as these experiments are optimized for the detection of Cherenkov light from high-energy muons (``muon tracks''). Another advantage of muon tracks over cascade events, such as the ones reported in \cite{Ishihara2013352}, is the higher directional information which allows for additional cuts in the search for point sources. Our photohadronic interaction cross
section includes several contributions apart from the $\Delta$-resonance, such as
higher resonances, direct production of pions ($t$-channel), and high-energy
processes leading to multiple pions. For the flavor mixing we still assume a
scenario with $\theta_{13} = 0$ even though Daya Bay, see \cite{An:2012eh}, and RENO, see \cite{Ahn:2012nd}, have ruled this out. However, it was already shown that the effect of different values of the
neutrino mixing angles inside the current uncertainty on the neutrino flux
prediction is comparably small, see \cite{Baerwald:2010fk}, and can be
neglected at this stage. In \Fig~\ref{fig:CygX-3neutrinos}, we plot the expected
muon neutrino (and antineutrino) flux (in $\giga\electronvolt \, \centi\meter\rpsquared
\, \reciprocal\second$) for Cyg~X-3 in this simple 
model of a MQ jet as a gray solid curve. The shape of the spectrum is similar to the ones predicted for GRB since we apply the same assumption of a neutrino flux originating from internal shocks. Hence, we can see the same splitting into a double peak from muon and pion decay plus an additional component from kaon decay, as in \cite{Baerwald:2010fk}. It is still significantly below the detection limit of IC59 during the time the flares were recorded (assumed exposure of 61 days, black solid curve), and if one extrapolates to the full IceCube detector and 5 yr of exposure (using the optimistic estimate that the level of the flux on average stays at the level of the flaring episodes), the neutrino flux would just start to be in the detectable range (black dashed curve). Note that here depicted limits are based on the solid-angle-averaged effective areas at final cut level of the time-integrated point-like source search from \cite{Abbasi:2010rd}. We use the band for $\delta = (30\degree,60\degree)$ from the left plot of \Fig~8 of said reference, which should be suitable for Cyg~X-3 ($\delta = 40\degree \, 57'$). Even though this effective area is already at the final cut level it still needs to be considered that the neutrino signal needs to be distinguished from the background of atmospheric neutrinos. There are, however, possibilities to improve the cuts, such as additional cuts in the timing as done in \cite{IceCube:2012sj} for Cyg~X-3 flares. Additionally, we have scaled the effective area up from the given values for the 40-string configuration to the 59-string (multiplied by $1.3$) and 86-string configurations (multiplied by $2.4$). Therefore, the actual effective areas for IC59 and IC86 point source searches may actually be slightly different, but were not publicly available during our work.
The gray shaded area represents the uncertainty on the flux due to the not directly measured emission radius of the X-rays. It represents the flux limits if the radius were one order of magnitude smaller or larger than the assumed $10^8 \, \centi\meter$. As can be seen from \Fig~\ref{fig:CygX-3neutrinos} especially a lower radius can lead to a significantly enhanced neutrino flux. This can be attributed to the way the photon and proton densities are normalized. As can be seen from \Eq~\eqref{equ:luminosity}, the calculated energy density is proportional to one over the square of the assumed radius $l$. Hence, the assumed photon density is significantly higher, and also the estimated magnetic field strength as well as the calculated proton density are increased. This increase is, however, only partially transferred to the neutrinos as the higher magnetic field also leads to higher losses of the secondary particles. Nonetheless, it is evident that the unknown emission radius is a major uncertainty on flux prediction.

Moreover, it can be seen that the predicted neutrino flux above $10^5 \, \giga\electronvolt$ is already one order lower than the peak value and the flux drops off even more above $\peta\electronvolt$ energies. Therefore, it should be unlikely that Cyg~X-3 is a possible source for the two unidentified $\peta\electronvolt$ cascade events in IceCube, see talk by \cite{Ishihara2013352} at Neutrino2012, as events at lower energies, \ie, between $10^4$ and $10^5 \, \giga\electronvolt$, are far more likely. Furthermore, the number of events can be calculated from the flux prediction and the detector parameters using the formula
\begin{equation}
	 N = \int \mathrm{d}E \, A^{\text{eff}}_\nu (E) \, t_{\text{exp}} \, \frac{\mathrm{d}N(E)}{\mathrm{d}E} 
\end{equation}
with $A^{\text{eff}}_\nu (E)$ being the energy-dependent effective area (of IceCube), $t_{\text{exp}}$ being the exposure, and $\mathrm{d}N(E)/\mathrm{d}E$ being the (muon) neutrino spectrum (on Earth, after flavor mixing). We obtain that we would expect $0.02$ events in IceCube for the time of the flares in 2009 June and July (59 strings) and a total of $0.84$ events over 5 yr in the full detector. Especially, this first result of no events in IC 59 is (in a sense) reassuring as it is consistent with the current IceCube data, which so far does not suggest any ultra-high energy (UHE) neutrinos apart from the atmospheric ones, see \cite{IceCube:2012sj}. 

\begin{figure}
	\centering
	\includegraphics[width=0.5\textwidth]{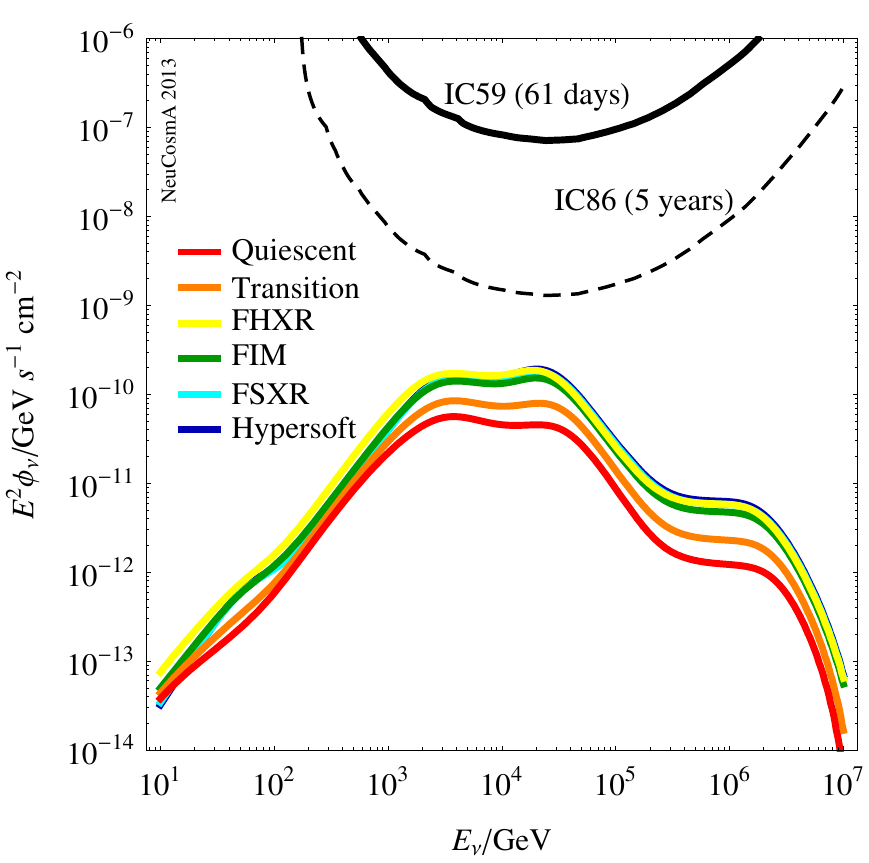} 
	\caption{Predictions for the six different X-ray flux states identified by \cite{Koljonen:2010wa}, see the legend inside figure. The calculations were done analogous to the hypersoft state for \Fig~\ref{fig:CygX-3neutrinos}, with a fixed collision radius $l = 10^8 \, \centi\meter$. As can be seen the variation between the highest predicted flux (for the hypersoft state) and the lowest predicted flux (for the quiescent state) is only of the order of two. Together with the differences in the predicted flux shape this gives a variation of more than three in the expected amount of neutrinos.}
	\label{fig:CygX-3fluxstates}
\end{figure}

Additionally, we compare the predicted result for the flaring episode (based on the hypersoft X-ray state) to the five other states identified by \cite{Koljonen:2010wa}. This is done to get an estimate of how much the dynamic of the source affects the prediction of the neutrino result. The calculation we use is the same as the one for the hypersoft state; however, we use the photon data of the other states to self-consistently calculate the energy densities in photon, protons, and the magnetic field. After having obtained these we again run our photohadronic interaction code to obtain the predicted neutrino flux here at Earth. As can be seen from \Fig~\ref{fig:CygX-3fluxstates}, the variation among the different states is only by a factor of about two, which is comparably small. The quiescent and the transition state give rise to a lower flux prediction than the hypersoft state, while the three other states, FHXR, FIM, and FSXR, are all roughly at the same level as the hypersoft state. There are, however, slight differences in the predicted flux shape which leads to a reduction by a factor of more than three when comparing the hypersoft state to the quiescent state. Therefore, even the lower flux states are predicted to be high enough to constrain elements of this simple MQ model within 5--10 yr of data taking with the full detector.

%\subsection*{Neutrinos from $\pi^0$ decay} \label{sec:pizero}
We now want to test how many neutrinos would be expected if the observed $\gamma$-ray emission by \textit{AGILE} was actually coming from the decay of photohadronically produced $\pi^0$ into photons. The photons from such decays would have to cascade down to lower energies and may lose a part of their energy during this process. However, here we only want to estimate the number of expected neutrinos for the case that the observed $\giga\electronvolt$ emission is due to $\pi^0$ decays. Hence, it is sufficient to assume energy conservation for the calculation. With this approach it is possible to set a lower bound on the expected amount of neutrinos. In case of energy losses during/after the cascading process an even higher amount of original $\pi^0$ (and consequently $\pi^+$) would be needed, and hence an even higher neutrino flux would be expected. If one assumes that 
\begin{equation}
	\int\limits^{10 \, \giga\electronvolt}_{10 \, \mega\electronvolt} \mathrm{d}E_\gamma \, E_\gamma \frac{\mathrm{d}N_\gamma (E_\gamma)}{\mathrm{d}E_\gamma} = \int\limits^{10^{10} \, \giga\electronvolt}_{0 \, \giga\electronvolt} \mathrm{d}E_\gamma \, E_\gamma \frac{\mathrm{d}N^{\text{HE}}_\gamma (E_\gamma)}{\mathrm{d}E_\gamma}
	\label{equ:pizeronorm}
\end{equation}
with $\mathrm{d}N_\gamma (E_\gamma)/\mathrm{d}E_\gamma$ being the observed photon spectrum from \cite{Bulgarelli:2011qk} while $\mathrm{d}N^{\text{HE}}_\gamma (E_\gamma)/\mathrm{d}E_\gamma$ is the resulting spectrum of high-energy photons before cascading to lower energies. The high-energy spectrum has been calculated numerically with the photohadronic interaction code, which also has been used for calculating the neutrino spectra. With this method we obtain that the amount of produced $\pi^0$ needed to match the observed $\gamma$-ray emission is about $2400$ times larger than our prediction for $\pi^0$ from the calculation for the hypersoft state in the previous section. As a consequence, the nominal amount of expected neutrino events would reach about $5.2$ events for the 61 days of flaring in 2009. 
Note that such a high amount of events would also be expected for the other flux states. Using the results of the other flux states and normalizing their respective results individually to the \textit{AGILE} data, we obtain expected amounts of $5.1$--$6.9$ neutrino events under the premise that the observed gamma rays are from $\pi^0$ decays. Here, the variation in the result is due to the different flux shapes and not the level of the actual observed photon data, as opposed to the flux predictions in \Fig~\ref{fig:CygX-3fluxstates}. Nonetheless, these few events would still need to be distinguished from the $\mathcal{O}(100)$ background events.\footnote{The amount of events is approximated from scaling the amount of events at final selection level from \Fig~4 of \cite{Abbasi:2010rd} down to 61 days (from 375.5).} If the used cuts achieve such a high precision, then it should be possible to rule out if the observed flares are due to the decay of $\pi^0$.

\section{Discussion} \label{sec:discussion}

In this paper, we have investigated the possibility that MQs may be the sources of high-energy neutrinos. In particular, we have estimated the neutrino flux expected from Cygnus~X-3. Starting from the X-ray data for the hypersoft state from \cite{Koljonen:2010wa}, we have calculated back to the particle densities of protons and photons inside the jet of Cyg~X-3 in the assumption of a simple geometrical model for the jet. The reason for choosing this data set was to use a data for a phase before a radio flare, as suggested by \cite{Bulgarelli:2011qk}. This state also fulfills the observed correlation of \textit{AGILE} $\gamma$-ray flares in 2009 June--July and 2009 November--2010 July with soft X-ray states and episodes of decreasing or non-detectable hard X-ray emission reported by \cite{Bulgarelli:2011qk}. However, in principle any other state could also be used for the calculations, as long as we are able to derive the particle densities inside the jet. 
We then used the densities to compute the expected neutrino emission from the jet using a numerical code which incorporates the full photohadronic interaction cross section, individual treatment of secondary particles (including losses), and flavor mixing of the neutrinos. The expected muon neutrino flux was then compared to the sensitivity of IceCube (59 strings) during the 61 days of flaring. The expected number of $0.02$ events is in concordance with the non-detection of any neutrinos from Cyg~X-3 during that period. Assuming an (optimistic) extended period of flaring and a full IC86 detector we expect about $0.84$ associated neutrino events in 5 yr of data taking. Moreover, the shape of the neutrino flux disfavors a detection at $\peta\electronvolt$ before seeing events at about $10 \, \tera\electronvolt$. Additionally, we compared the predicted neutrino flux levels for the flares to the other states identified by \cite{Koljonen:2010wa} to incorporate the dynamic of the source. The resulting change in neutrino flux levels was only of $\mathcal{O}(1)$ and only reduces the expected amount of neutrinos mildly. Nevertheless, these predictions are still subject to some uncertainties due to the not directly measured emission radius as well as the used extrapolated effective area of IceCube.
Therefore, even slightly lower amounts of observed neutrinos may not directly contradict the basic model of MQs accelerating protons. Still, in the best case it should be possible to see some events from Cyg~X-3 with several years of data taking with the full IceCube detector.
Moreover, we also tested the hypothesis that the observed $\gamma$-ray emission is due to the decay of $\pi^0$ from photohadronic interactions into photons. We would like to add that the $\pi^0$ production of the gamma-ray flux is not likely if ``the simplified jet model'' is the way gamma rays are produced. In any case to test this hypothesis we compared the integrated energies in the observed photons detected by \textit{AGILE} to the energy in photons from $\pi^0$ decay. In the $\Delta$-resonance approximation, the production of $\pi^0$ is directly connected to the production of $\pi^+$, and this concept in principle does not change even for a more detailed particle physics treatment of the interactions. Using these calculations, we obtained that the amount of energy in $\pi^0$ needed to explain the \textit{AGILE} observations would have been so high that the number of expected neutrino events would reach about five events during the 61 days of flaring. There have, however, been no reported neutrino events in IceCube which could be associated with Cyg~X-3 so far, see \cite{IceCube:2012sj}. We can therefore start to rule out that the observed $\gamma$-ray emission is due to the decay of $\pi^0$ from photohadronic interactions by combining the photon and neutrino data in the coming months. Especially, the point source analysis of IC59 would be of great interest in this regard.

%\acknowledgments

\section*{Acknowledgements}

We thank Francis Halzen, Naoko Kurahashi Neilson, Giovanni Piano, Marco Tavani, Eli Waxman, Nathan Whitehorn, and Walter Winter for the helpful discussions and
comments. DG thanks the Weizmann institute where part of this research has been
carried out. PB thanks the Weizmann Institute for hospitality and support during his stay.
PB acknowledges support from the GRK1147
``Theoretical Astrophysics and Particle Physics'' and the ``Helmholtz Alliance
for Astroparticle Physics HAP'', funded by the Initiative and Networking fund of the Helmholtz association.

%\bibliographystyle{apj}
%\bibliography{references}

\end{document}